# Computational Design of Nanoclusters by Property-Based Genetic Algorithms: Tuning the Electronic Properties of $(TiO_2)_n$ Clusters

Saswata Bhattacharya[1], Benjamin H. Sonin[2], Christopher J. Jumonville[2], Luca M. Ghiringhelli,[*,1] and Noa Marom,[*,2]

[1]*Fritz-Haber-Institut der Max-Planck-Gesellschaft, Faradayweg 4-6, 14195, Berlin, Germany*
[2]*Physics and Engineering Physics, Tulane University, New Orleans, Louisiana 70118, USA*

In order to design clusters with desired properties, we have implemented a suite of genetic algorithms tailored to optimize for low total energy, high vertical electron affinity (VEA), and low vertical ionization potential (VIP). Applied to $(TiO_2)_n$ clusters, the property-based optimization reveals the underlying structure-property relations and the structural features that may serve as active sites for catalysis. High VEA and low VIP are correlated with the presence of several dangling-O atoms and their proximity, respectively. We show that the electronic properties of $(TiO_2)_n$ up to $n=20$ correlate more strongly with the presence of these structural features than with size.

The rapidly growing field of nanocatalysis [1-4] harnesses the tunability of nano-structured materials to increase the activity and selectivity of catalysts, while lowering their cost by reducing the amount of rare elements. The properties of nanocatalysts depend strongly on the size, composition, and local structure of their constituents. Nanocatalysts often comprise metal or metal-oxide clusters dispersed on a high surface area support, owing to the enhanced reactivity of clusters compared to bulk materials [5-13]. In addition, clusters may serve as models to reveal the structure and function of catalytically active sites on surfaces [14-16].

The role of first principles simulations has been expanding from elucidating the active sites and reaction mechanisms involved in catalytic processes to computational screening and design of potential catalysts [17-21]. Efforts to computationally design catalysts have largely focused on solid alloys and surfaces. At the same time, the vast majority of computational studies of clusters have focused on searching for their global minimum structures [10, 22-25]. However, the most stable structures of atomic clusters are not necessarily optimal for catalysis. Rather, the presence of active sites has been shown to be a key factor in catalysis by clusters [26-31].

Our quest to computationally design atomic clusters for applications in nanocatalysis embarks from the hypothesis that clusters possessing a high vertical electron affinity (VEA) or a low vertical ionization potential (VIP) may be more promising candidate catalysts because they would be more chemically reactive (as they would accept or donate an electron more readily). We further assume that these electronic properties are correlated with the presence of potentially active sites [32]. Based on these premises, we have implemented a suite of three massively parallel cascade genetic algorithms (GAs). The first is the energy-based GA (EGA), described in detail in [33, 34]. The second is tailored to search for clusters with a high VEA (VEA-GA) and the third is tailored to search for a low VIP (VIP-GA). Here, we apply them to the case of $(TiO_2)_n$ clusters with n=2-10,15,20.

Analysis of the optimal structures found by the property-based GAs reveals the underlying structure-property relations. We show that the electronic properties of $(TiO_2)_n$ clusters depend more strongly on structure than on size. In particular, we find a clear correlation between certain structural motifs and the magnitude of the VEA, VIP, and fundamental gap. Contrary to common belief, we do not find a clear size trend. We attribute this to the fact that particular structural features related to high VEA and low VIP are less favorable energetically and therefore less likely to be found for larger clusters.

GAs have been used extensively for finding the energy global minimum (GM) of crystalline solids [35-40] and clusters [22-25, 41]. The strategy of a GA is to perform global optimization by simulating an evolutionary process [22-24, 42]. First, local optimizations are performed for an initial population of randomly generated structures. The scalar descriptor (or combination of descriptors) being optimized is mapped onto a fitness function. The structures with the highest fitness are assigned a higher probability for mating. In the mating step a crossover operator creates a child structure by combining fragments (or structural "genes") of two parent structures. The child structure is locally optimized and added to a common pool if determined to be different than the existing structures. The cycle of local optimization, fitness evaluation, and generation of new structures is repeated until convergence is achieved. Advantageously, the fitness function may be based on any property of interest, not necessarily the total or free energy [22, 43-47]. We rely on this to tailor GAs to explore the configuration space of desired electronic properties.

Within our cascade GA approach [33, 34] successive steps employ an increasingly accurate level of theory. Structural information is passed between steps of the cascade and some structures are filtered out. This considerably increases the efficiency of the GA and reduces the time to solution.

Initially, the reactive force field ReaxFF [48-50] is used for an exhaustive GA pre-screening of possible structures. Structures found within an energy window of 4 eV from the energy GM are then transferred to a density-functional theory (DFT) based GA. DFT calculations are performed with the all-electron numeric atom-centered orbitals code FHI-aims [51, 52]. The cascade flow proceeds as follows: In the first step local optimizations are performed with the Perdew-Burke-Ernzerhof (PBE) [53, 54] functional, using *lower-level* settings [55]. Structures that are already in the pool or outside an energy window of 2 eV above the running GM are rejected [56]. The remaining structures are passed to the second step of the cascade, where their energies are calculated with the PBE-based hybrid functional (PBE0) [57] to evaluate their fitness [58]. For the electronic property based GAs, the VEA/VIP are evaluated based on the total energy difference between the neutral species and the anion/cation. After the GA cycle converges the 50 fittest isomers (for each cluster size) are post-processed at the *higher-level* settings [55]. The structures are first re-optimized using PBE. Then, PBE0 is used for the final single point total energy and VEA/VIP evaluations. We have verified that among isomers the *hierarchy* of the PBE0-level electronic properties (total energy, VEA, VIP) is largely conserved upon switching from *lower-level* to *higher-level* settings. Since the hierarchy of the optimized quantity is more important than the absolute values, this enables a considerable reduction of the computational effort by running the potential energy surface scan with *lower level* settings. A complete account of the GA implementation and validation is provided in the SI, including details on how crossover, mutation, and similarity checks are performed and how convergence is determined [59].

The three GAs ran independently starting from the initial pool generated by the force field based GA until convergence was achieved. In Figure 1 their performance is compared in terms of the ability to find structures that optimize different target properties. Histograms of the number of structures found for each property value by the respective GA are shown for $(TiO_2)_n$ clusters at selected sizes. The data shown here were obtained using the PBE0 functional with the *lower-level* settings, reflecting the level of theory used to evaluate the fitness function in the final step of the cascade scheme.

For small clusters (n=2,3) the three algorithms find the same isomers because the configuration space is small and there are only a few structures in the search window. Starting from n=4, it is evident that the three GAs explore different regions of the configuration space. The centers of the VEA-GA histograms in panel a are clearly shifted to higher VEAs, compared to the EGA histograms and there are more structures in the tail regions with particularly high VEA. Similarly, the VIP-GA histograms in panel b are shifted to lower VIPs. From the energy histograms in panel c, it is evident that the VEA-GA and VIP-GA systematically explore higher energy regions of the configuration space than the EGA. Hence, they efficiently find more structures with the desired electronic properties than the EGA [60].

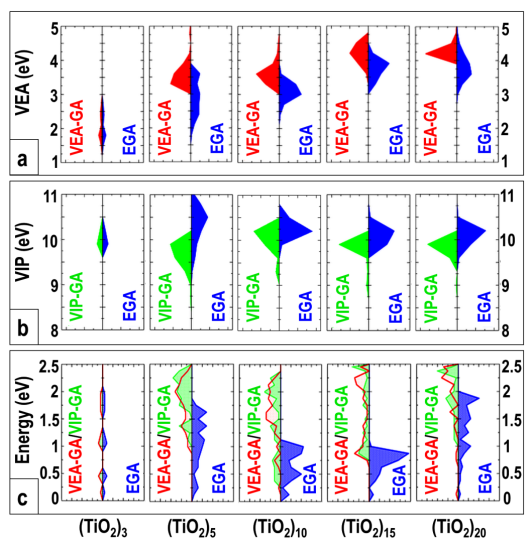

**Figure 1.** Comparison of the three GAs in terms of their ability to find structures with the target properties of (a) high VEA (b) low energy. Histograms represent the number of isomers found by the respective GA for $(TiO_2)_n$ clusters with n=3,5,10,15,20.

Analysis of the structures found by the property-based GAs reveals the structural features correlated with a high VEA and a low VIP. Figure 2a shows the Ti-O pair correlation functions (PCFs) of the ten best $(TiO_2)_5$ clusters found by the VEA-GA and the EGA. A peak at 1.65 Å is clearly more prominent in the VEA-GA PCF, compared to the EGA PCF. This is observed for all cluster sizes (see SI). This peak corresponds to the bond length of dangling-O atoms, shown in Figure 2b. Figure 2c shows the average and maximum VEA (upper panel) and number of dangling-O atoms (lower panel) for the same sample of isomers as for the PCFs. The data shown from here on (including in Figures 3, 4) correspond to the *higher-level* settings of the final post-processing stage.

For the smallest clusters with n=2,3 the graphs overlap because both algorithms find the same isomers. Starting from n=4, the graphs diverge. The high VEA isomers clearly have a larger number of dangling-O atoms than the low energy isomers. Therefore, a correlation can be drawn between the number of dangling-O atoms and a high VEA. Many of the high-VEA isomers with n≥4 have 3-4 dangling-O atoms. The number of dangling-O atoms decreases for high-VEA clusters with n≥10 because this structural motif becomes increasingly unfavorable energetically with size. Many of the high VEA clusters also have the tri-coordinated Ti site, reported in [32], as shown in the SI. Thus, the VEA-GA has revealed a new structure-property relation.

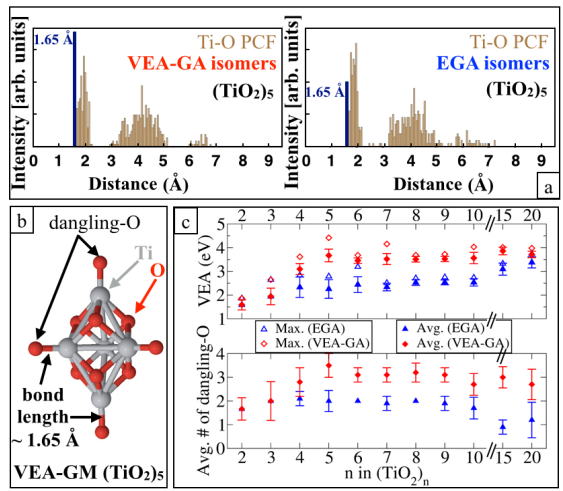

**Figure 2.** a) Cumulative Ti-O PCFs of the ten best $(TiO_2)_5$ isomers found by the VEA-GA (top) vs. the EGA (bottom); b) Visualization of the $(TiO_2)_5$ isomer with the highest VEA, showing its four dangling O atoms; c) The average and maximum VEA (top) and number of dangling O atoms (bottom) of the 10 isomers with the highest VEA vs. the 10 isomers with the lowest energy for all cluster sizes.

Visual inspection of the structures found by the VIP-GA reveals that they tend to have two dandling-O atoms in proximity to each other. The corresponding descriptor is the *bond connectivity* (the number of Ti-O bonds along the shortest path between two Ti atoms attached to dangling-O atoms, see Fig. 3a). Figure 3b shows the average and minimum VIP (upper panel) and the bond connectivity (lower panel) for the 10 isomers with the lowest VIP (found by the VIP-GA) vs. the 10 isomers with the lowest energy (found by the EGA) for all cluster sizes. Starting from n=4, the bond connectivity of the low-VIP isomers is consistently lower than that of the low-energy isomers. For n=4-9, the lowest VIP isomers have a bond connectivity of 2. For larger clusters (n≥10), the bond connectivity is higher. The reason is that energetically unfavorable under-coordinated O atoms are less likely to occur in larger clusters and if they do occur they are less likely to be in proximity to each other. This observation is supported by the fact that the EGA does not produce any isomers with more than one dangling-O for n=15. The VIP-GA does find structures with a low VIP and a bond connectivity of 2 for n≥10, however they fall outside of the 2 eV energy window and are therefore not shown in Figure 3.

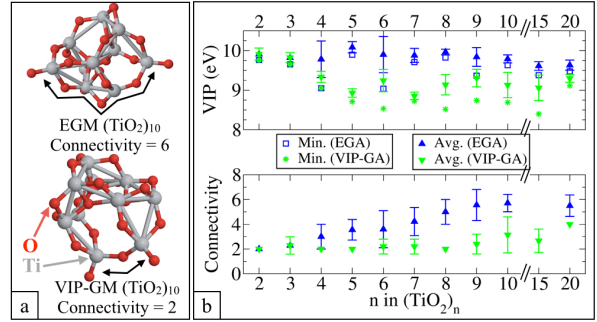

**Figure 3.** a) Bond connectivity of the energy global minimum (EGM) vs. the VIP-GM for $(TiO_2)_{10}$; b) The average and minimum VIP (top) and bond connectivity between dangling-O atoms (bottom) of the 10 isomers with the lowest VEA vs. the 10 isomers with the lowest energy for all cluster sizes.

Owing to quantum confinement effects, one would expect size trends in the electronic properties of clusters, whereby the VEA increases, the VIP decreases, and the gap narrows with size. The expected size trends are not readily apparent in Figures 2c and 3b. The VEA increases fast for the smallest isomers and then stabilizes and becomes almost constant. The VIP decreases for the smallest clusters and then fluctuates. We further investigate whether there is a size trend in the fundamental gap (VIP–VEA) of the structures found in the three searches.

Figure 4a shows the VIP vs. VEA for the best 10 structures found in each search for all cluster sizes. The loci of constant fundamental gap are indicated by diagonal lines. The results of the EGA, VEA-GA, and VIP-GA are clustered in different regions of the graph. The clusters found by the VEA-GA and VIP-GA tend to have gaps of 5.5-6.5 eV while the clusters found by the EGA tend to have gaps of 7-8 eV. Considering that a narrow gap requires a combination of a high VEA and a low VIP, it is not surprising that the VEA-GA and VIP-GA generally find structures with smaller gaps than the EGA [61].

What is more unexpected is that the clusters with the smallest gaps are not necessarily the largest ones. The three clusters with gaps below 5 eV have n=6,15, and 10. Between 5 to 5.5 eV we find clusters with n=5,7,8,9,10,15, and 20. Figure 4b shows the average and minimum gap of the 10 best structures found in each search for all cluster sizes. A clear trend of narrowing gap with increasing size is seen only for the smallest clusters. For the larger clusters the EGA isomers show a weak trend of gap narrowing with size. The isomers found by the VEA-GA and VIP-GA exhibit significant fluctuations of the minimum and average gap and no clear size trend is visible.

The absence of the expected size trends may be explained by the structural features associated with a high VEA, a low VIP, and a narrow gap. These features become less energetically favorable with increasing size because they involve a number of dangling-O atoms, some of which are in proximity to each other. Therefore, they are less likely to appear in larger clusters. We have thus demonstrated that the electronic properties of $TiO_2$ clusters with up to 20 stoichiometric units correlate more strongly with the presence of specific structural features than with the cluster size. For larger clusters we expect to eventually reach a size regime where isomers possessing multiple dangling-O atoms completely disappear from the search window and the electronic properties correlate more strongly with size. Indeed, such size trends have been reported for the IP, EA, and gap of bulk-like rutile $TiO_2$ nanocrystals without any dangling-O sites [62].

In summary, we have implemented a suite of three cascade genetic algorithms tailored to optimize cluster structures for low total energy (EGA), high vertical electron affinity (VEA-GA), and low vertical ionization potential (VIP-GA). Analysis of the structures found by the VEA-GA and the VIP-GA vs. the EGA reveals the following structure-property relations: A high VEA is correlated with a number of dangling-O atoms (typically 3-4) and a low VIP is correlated with low bond connectivity (typically 2) between two dangling-O atoms. These structure-property relations explain the absence of the expected size trends. Smaller clusters may have a higher VEA, a lower VIP, and a narrower gap than larger clusters because the structural features associated with these properties become less favorable with increasing size.

We further suggest that the presence of dangling-O atoms on $TiO_2$ clusters or surfaces may be associated with enhanced catalytic activity and that these O atoms may serve as the active sites. Our findings hint at a new approach to the computational design of cluster-based nanocatalysts, employing property-based GAs. These broadly applicable algorithms may be modified to search for any desired electronic property (that can be mapped into a fitness function) and may be extended to clusters of varying composition as well as cluster/support systems. The process of optimization for a target property, associated with enhanced reactivity, reveals the underlying structure-property relations and the structural features that may serve as active sites for catalysis. This may provide valuable physical insight and design rules for better nanocatalysts.

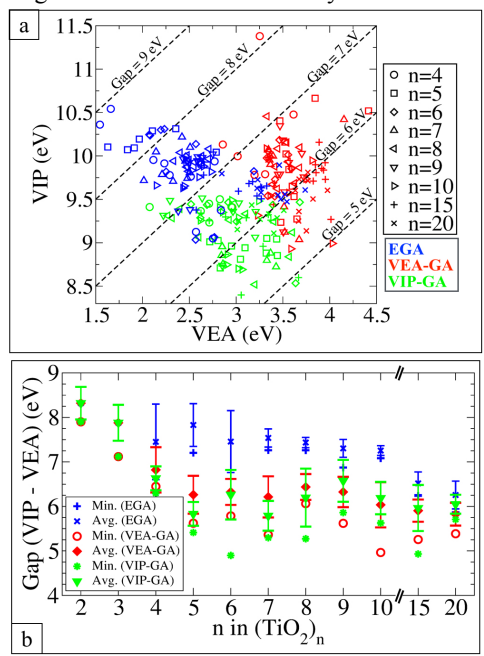

**Figure 4.** a) VIP vs. VEA for the best 10 structures found in each search for all cluster sizes. The loci of constant fundamental gap are indicated by diagonal lines. b) The average and minimum gap of the 10 best structures found in each search for all cluster sizes.

We thank Kristen Fichthorn and Muralikrishna Raju from Penn State for sharing the reaxFF parameters they used for $TiO_2$ surfaces. We thank Matthias Scheffler and Sergey Levchenko from FHI-Berlin for helpful discussions and a critical reading of the manuscript. Work at Tulane University was supported by the Louisiana Alliance for Simulation-Guided Materials Applications (LA-SiGMA), funded by the National Science Foundation (NSF) award number #EPS-1003897. Computer time was provided by the Argonne Leadership Computing Facility (ALCF) at Argonne National Laboratory, which is supported by the Office of Science of the U.S. Department of Energy under contract DE-AC02-06CH11357 and by the Rechenzentrum Garching (RZG) of the Max-Planck Gesellschaft differences between isomers. This is reflected in Figure 1c.

# Computational Design of Nanoclusters by Property-Based Genetic Algorithms: Tuning the Electronic Properties of $(TiO_2)_n$ Clusters

Saswata Bhattacharya[1], Benjamin H. Sonin[2], Christopher J. Jumonville[2],
Luca M. Ghiringhelli[1*], Noa Marom[2†]

[1]*Fritz-Haber-Institut der Max-Planck-Gesellschaft, Faradayweg 4-6, D-14195 Berlin, Germany and*
[2]*Physics and Engineering Physics, Tulane University, New Orleans, Louisiana 70118, USA*

(Dated: January 22, 2015)

# Supplemental Material

## Contents

---

\* corresponding author (LMG): ghiringhelli@fhi-berlin.mpg.de
† corresponding author (NM): nmarom@tulane.edu

# I. COMPUTATIONAL DETAILS:

## A. Implementation of *propery based cascade* Genetic Algorithm:

We investigate whether electronic properties other than total energy are associated with specific structural features and with the presence of active sites. This necessitates unbiased sampling of the potential energy surface (PES) of atomic clusters at the nano- and subnano-scale, we have implemented a suite of three massively-parallel cascade genetic algorithms (GA) schemes. The first scheme, as described in detail in Ref. [1, 2], is our recent implementation of a cascade genetic algorithm, where the minimized quantity is the total energy (Energy-GA, EGA), which is the conventional minimized quantity for atomic structures. For the second and third schemes, the fitness function is defined such that the largest VEA (vertical electron affinity) - viz., lowest VIP (vertical ionization potential) - is searched. As a test system, we have chosen $(TiO_2)_n$ clusters with n=2-10,15,20.

The strategy of a GA algorithm is to optimize structures by simulating an evolutionary process. First, local optimizations are performed for a randomly generated initial population of structures. The fitness of those structure is then evaluated with respect to a (scalar) property to be globally optimized. Here the property is total energy, VEA, VIP, in EGA, VEA-GA, VIP-GA schemes, respectively. Then two parent structures are selected using a roulette-wheel selection criterion [3] where the larger is the fitness value, the higher is the probability of getting selected. A trial structure is created by combining two fragments of the selected structures. In this mating step, the crossover operator takes care of combining two such fragments from the parent clusters. It is implemented as a modified version of the cut-and-splice crossover operator of Deaven and Ho. [4] Crossover generates a child, and following that mutation is introduced. Amongst many different possible mutation schemes, here we have only adopted a random translation between the two halves of the parent clusters, if atoms coming from the two different parents are too close after splicing of the two halves. After that the child structure is locally optimized (either using a forcefield or DFT with semi-local functional) and a new structure is obtained. To decide whether the newly found structure was already seen previously during the GA scan, after the local optimization we i) compare the total energy of the new structure with that of all the others seen before and ii) use a criterion based on the distances between all the atoms' pairs. We do this by constructing a coarse grained pair-correlation function (PCF) of the clusters, consisting of $n$ bins conveniently spaced. (see details as in [2]). If the newly found structure was unseen before, we add this in the existing pool and the cycle is repeated until convergence is achieved. It should be noted here for any search performed in a multidimensional configuration space, it is generally almost impossible to guarantee that the global optimum (GO) is found. Therefore, we use an operative definition of convergence (see also Ref. [2]). After a transient initial number of GA cycles, in which the fittest structure (the running GO) changes rapidly, we reach a regime when only rarely a new running GO is found. In this regime, we always keep track of the total number of GA cycles needed to find the running GO and we make sure to run at least as many cycles further, before concluding that a new GO is unlikely be found and the running GO is confirmed as the searched GO structure. It should be noted here that to facilitate transitions between basins during the GA search, the population diversity is always maintained by occasionally selecting some unfit structure for mating. [2]

The GA algorithm is massively parallelized. The operation of selecting two structures from a genetic pool for the mating and the subsequent local optimization of the child, can be performed absolutely independently at any moment. Thus the scaling with respect to the replicas is almost linear, because the replicas remain independent for the most of the time and only at the beginning and at the end of each local optimization, the information is exchanged among the replicas. Therefore, first level of parallelization is performed within the FHI-aims code [5], by means of the MPI environment. [6] The second level of parallization is script based. In total $n \times p$ number of cores is equally divided into $n$ replicas, where all the replicas perform independently a cycle of local optimizations each using p cores. The algorithm is, therefore, suitable for a very efficient parallelization. This second level of parallelization scales practically linearly with the number of cores, as there is no idling time between subsequent local-optimization cycles (a replica does not need to wait for the other replica before launching the next local optimization).

It's important to note that, the GA has the important advantage that the fitness function can be optimized to be any property of our interest, and not necessarily to be the conventionally used total or free energy.

We adopt our recent implementation (namely, cascade GA approach as in Ref. [1, 2]) to scan the configuration space of $(TiO_2)_n$ clusters. The term "cascade" refers to a multi-stepped procedure where successive steps employ higher level of theory and each of the next level takes information obtained at the immediate lower level. Therefore, an exhaustive GA pre-scanning of large number of possible structures is done by means of a computationally inexpensive, albeit rather sophisticated, reactive force field (ReaxFF). [7, 8] Structures found within a certain energy window from the energy global minimum (GM) are used as initial GA pool to start a fully quantum mechanical GA based on density functional theory (DFT). The energy window depends on the size of the system (i.e.; number of atoms) and is usually taken as sufficiently large ( 4 eV) to generate a diverse initial pool for the DFT-based GA. We choose this window in such a way that the initial pool for DFT-GA is comprising a few tens (for a small system of $\leq$ 10 atoms) to several hundred structures (for a medium to large system $\geq$ 40 atoms). As shown in Ref. [2], it is generally unjustified to use only the low-energy structures found via a force field based GA as representative of the low-energy DFT structures, even if subsequent geometry optimization is performed using DFT. We strongly recommend, a complete DFT-based GA optimization should be performed with the structures generated by the force field based GA.

As the cascade starts to the DFT level, the balance between accuracy and efficiency is extremely important. The probability of selection for mating depends on the fitness function and the fitness function depends on the accuracy of the electronic properties evaluated at the DFT-level. It is, therefore, important to accurately calculate the electronic properties used to define the fitness function. It has been shown that the energy hierarchy of structures and electronic properties, such as the VEA/VIP, are sensitive to the employed DFT functional (or higher level theoretical approach). [2, 9] However, it would be computationally impractical to perform the whole DFT-GA scan at the highest level of theory. In view of this, the lower-accuracy steps of the cascade algorithm becomes important that provides considerable prescreening of structures, thereby saving a lot of computer time, without compromising the accuracy of the fitness function evaluation in the more promising regions of the PES.

All DFT calculations are performed with the all electron numeric atom-centered orbitals (NAO) code FHI-aims. [5] At the DFT level, the cascade flow continues with two sub-steps. Initial local optimizations are performed with the Perdew-Burke-Ernzerhof (PBE) [10, 11] functional using *lower-level settings*[16]. Structures that are similar to structures already in the pool or outside an energy window of 2 eV above the running GM are not considered for the next cascade step. However, the new structures found within 2 eV energy window from the running GM, are taken to the next level of cascade where single point total energy of the cluster is estimated using the PBE-based hybrid functional (PBE0). [12] This latter energy is used to calculate the fitness function of that cluster and is included in the GA-pool. For the electronic property based GAs at this stage the vertical VEA/VIP are evaluated from the total energy difference between the neutral species and the anion/cation, using PBE0 functional. These VEA/VIP values are used to estimate the fitness function. Based on Ref. [9], isomers with energy as high as 1.25 eV above the GM may have been observed in experiments, in which $TiO_2$ clusters were formed by laser vaporisation. Restricting the search window to 2 eV would not eliminate any isomer that can conceivably form in such experiments. Note that this 2 eV cut-off window is set at the PBE *lower-level settings* (at the DFT-cascade: step1) and after single point total energy calculation using PBE0 at the next cascade level (DFT-cascade: step2), we found that this window gets changed by 2.5 eV from the EGM. This is shown in Fig. 1 (bottom panel) of the main paper. In this figure, the relative energies, VEAs, and VIPs were obtained at the lower-level settings using PBE0 functional and this reflects the level of theory at which the fitness function is evaluated. The VEAs, VIPs, and gaps shown in Fig. 2-4 (main paper) were obtained at the higher-level settings using PBE0, reflecting the final post-processing step as explained in the section I(c).

### B. Validation and comparison of DFT-GA at *lower* and *higher-level* settings:

It should be noted here that in Ref. [2] we have employed *higher-level* settings at the final cascade step of DFT-GA scan, after a sub-step performed at *lower-level* settings. On the contrary, in this present work we have performed the entire DFT-GA scan only at the *lower-level* settings and no *higher-level* settings are employed inside cascade steps of DFT-GA.[17] This is because in the present system the *hierarchy* of electronic properties of different isomers evaluated at the *lower-level* settings follows exactly similar *pattern* as that of evaluated at *higher-level* settings [see Fig. 1]. The *lower-level* settings give the correct hierarchy of the VEA/VIP, although the absolute values may change by 0.1-0.5 eV upon switching to the *higher-level* settings. We note that, for the GA, the *hierarchy* of the globally optimized quantity is more important than the absolute values. Moreover, we have verified for $(TiO_2)_4$ as a test case, that running the GA at the *higher-level* settings yields the same set of low-energy isomers as running the GA at the *lower-level* settings. Therefore, this enables us to make a considerable reduction of the computational effort by doing the entire DFT-GA at the *lower-level* settings. Obviously, if that would not have been the case [e.g. the system described in Ref. [1, 2]], we would have needed to employ *higher-level* settings inside the DFT-GA scan as explained in details in Ref. [2].

### C. Post-process of the best fitted DFT-GA isomers using *higher-level* settings:

We have however noticed that the value of the electronic properties using *higher-level* settings differ by 0.1-0.5 eV as compared to the same found by *lower-level* settings. Therefore, in order to accurately calculate the VEA/VIP values of all different isomers, we have first optimized the structure using PBE (*higher-level* settings) and then two single point PBE0 calculation (*higher-level* settings) are performed at charge=0 and charge=-1/+1 (anion/cation). The VEA/VIP is therefore evaluated as the difference in total-energy between the neutral species and the anion/cation. We have named this methodological approach as PBE0 delta-SCF as explained in section I(d).

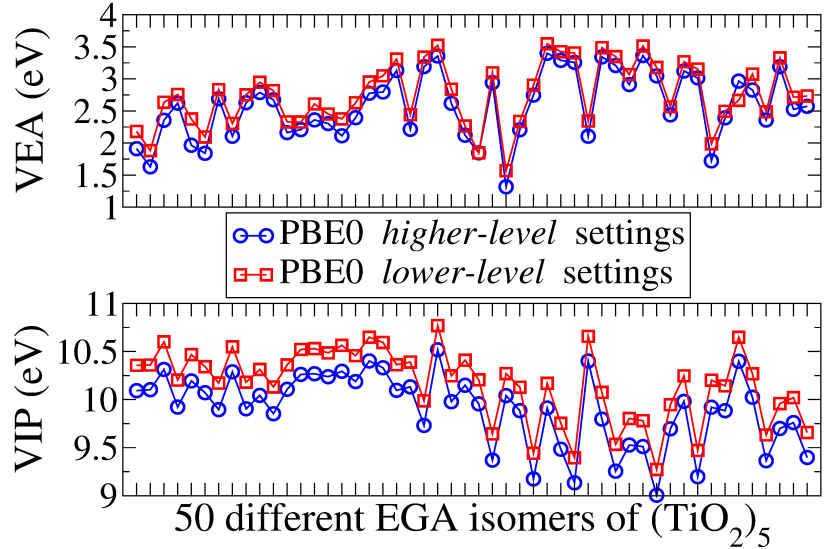

FIG. 1: *Hierarchy* of electronic properties (namely VEA, VIP) of $(TiO_2)_5$ EGM isomers (completely arbitrarily chosen) evaluated at *lower-level* and *higher-level* settings.

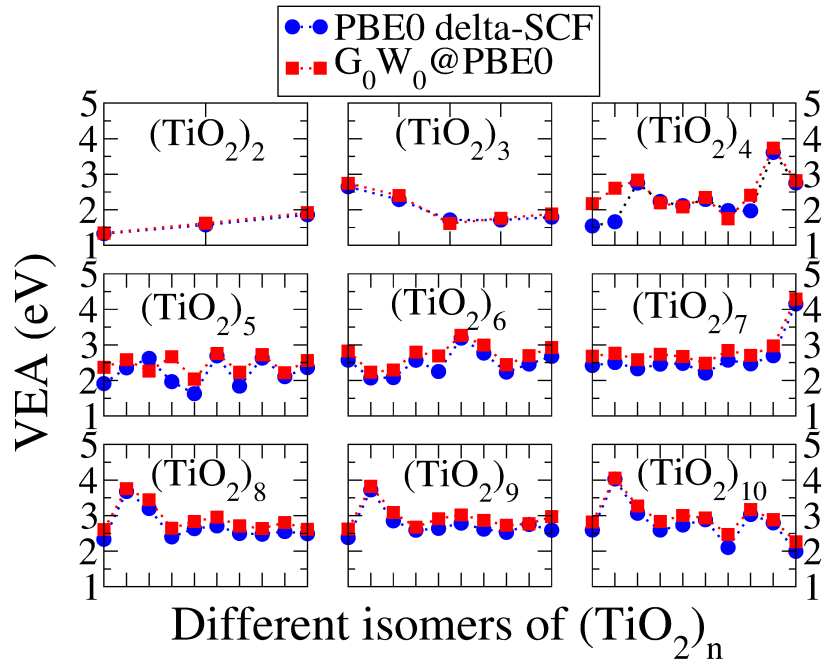

FIG. 2: Comparison of VEA values using $G_0W_0@PBE0$ vs PBE0 delta-SCF of randomly selected 10 different $(TiO_2)_n$ isomers.

### D. Validation of PBE0 delta-SCF energetics w.r.t $G_0W_0$@PBE0:

The VEA/VIP obtained from total energy differences with PBE0 at the *higher-level* settings are in good agreement with more accurate calculations based on many-body perturbation theory within the GW approximation, used in Ref. [9]. In addition, it has been shown in Ref. [13] that IPs and EAs calculated from DFT total energy differences exhibit similar size trends to GW calculations for bulk-like rutile $TiO_2$ nanocrystals. Thus, in order to understand the accuracy of VEA/VIP values as evaluated from PBE0 delta-SCF, we have randomly selected several isomers at all different sizes and compared with the VEA values as evaluated from many-body perturbation theory (e.g. $G_0W_0$@PBE0).

The absolute VEA values using $G_0W_0$@PBE0 are not always exactly matching with PBE0 delta-SCF values (with an estimated mean average error of  0.2 eV), but in all different sizes PBE0 delta-SCF qualitatively predicts similar energy profile as of $G_0W_0$@PBE0. It's clear that at all different sizes, PBE0 delta-SCF always provides the correct trends [see Fig. 2]. Note that the $G_0W_0$ calculations are performed using a large "tier4" basis set, while PBE0 delta-SCF calculations are done using "tier2" basis set. This validates the computational efficiency of our PES-search methodology.

## II. NO. OF DANGLING-O VS NO. OF $Ti^{3+}$ IN $(TiO_2)_n$ CLUSTERS:

It has been shown before that localization of the lowest unoccupied molecular orbital (LUMO) on a tri-coordinated Ti atom (namely, $Ti^{3+}$ site) is often associated with a high vertical electron affinity (VEA). [9] But we found many structures without $Ti^{3+}$ site that are having high VEA [see Fig. 3 below].

Therefore, we have given a comparative structural analysis of number of $Ti^{3+}$ site vs number of dangling-O in the best 10 isomers found from EGA-search and VEA-GA search [see Fig. 4]. It's true that many isomers with high VEA are having this $Ti^{3+}$ site especially at small cluster size, but with increasing size ($n$) of the cluster, many high-VEA isomers do not show this feature. In contrast, the correlation between having high VEA and increasing number of dangling-O in the structure holds at all cluster size ($n$) we have investigated.

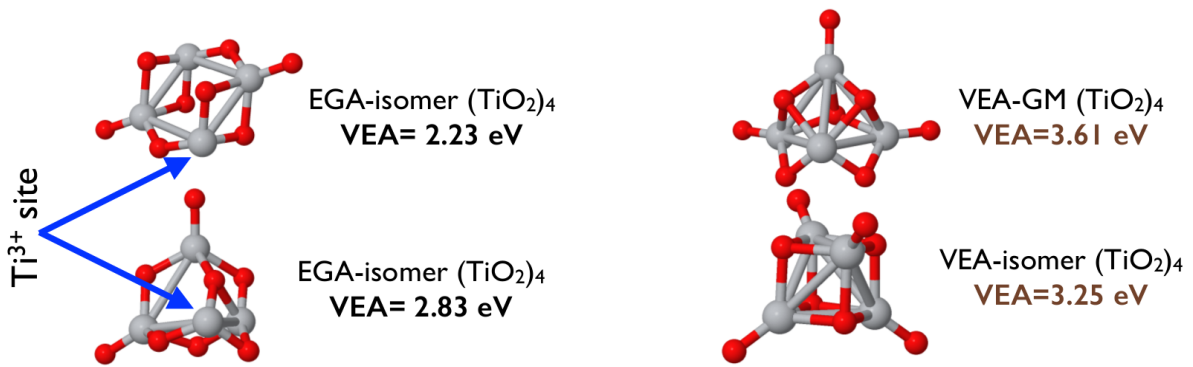

FIG. 3: The structures at the left-hand side have the $Ti^{3+}$ site(s), while this feature is absent in the structures as in the right. However, the VEA of the later is larger than the former. The structures at the right-hand side have a larger number of dangling-O's, compared to the structures at the left-hand side. See more comparisons in Fig. 6.

# no. of dangling-O vs no. of Ti³⁺ in (TiO₂)ₙ clusters

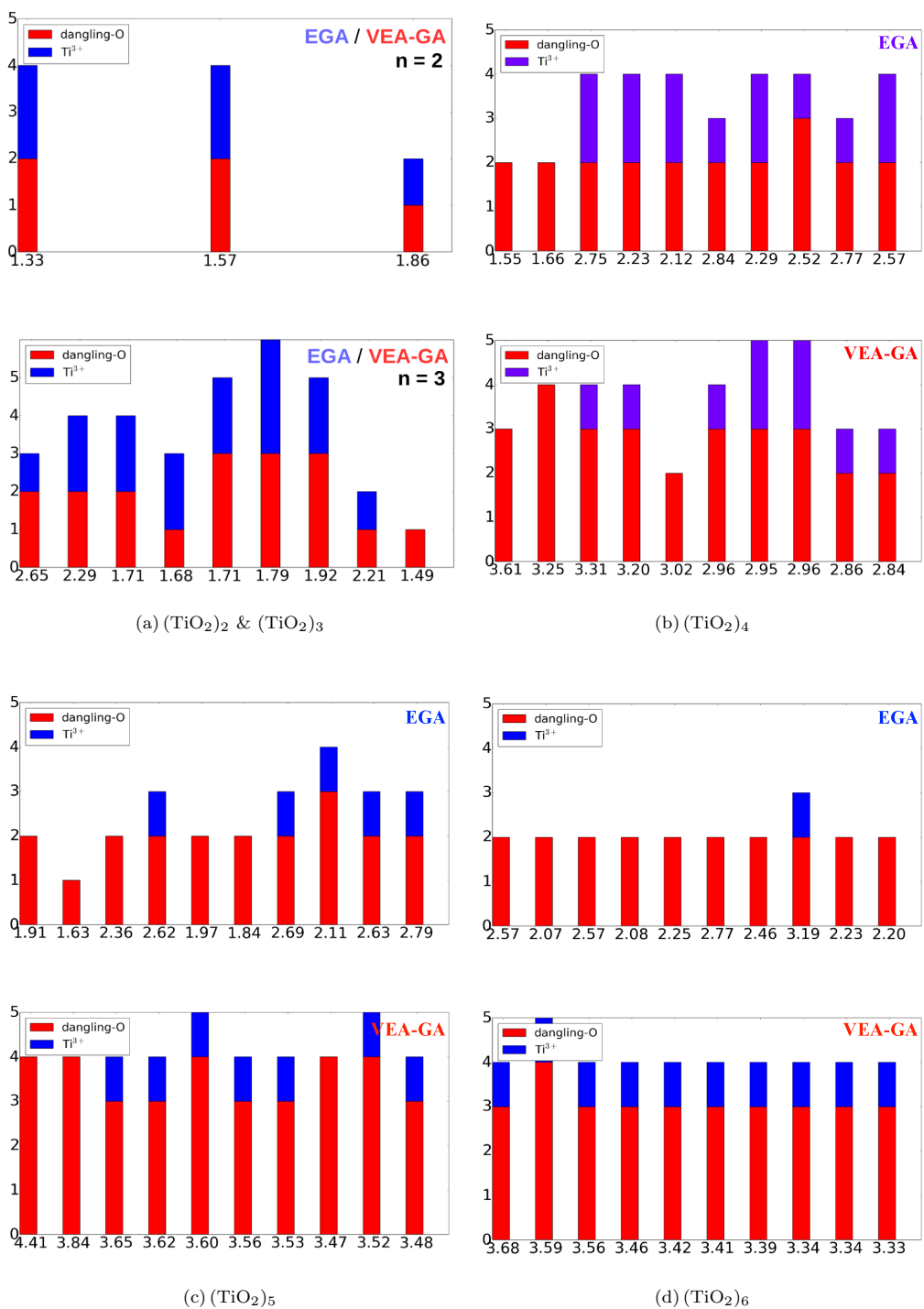

(a) (TiO₂)₂ & (TiO₂)₃

(b) (TiO₂)₄

(c) (TiO₂)₅

(d) (TiO₂)₆

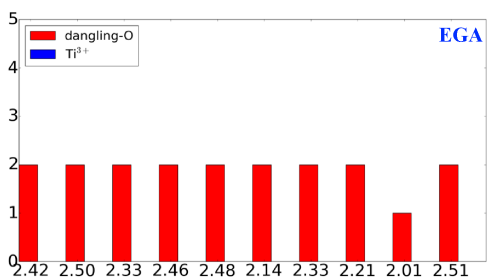

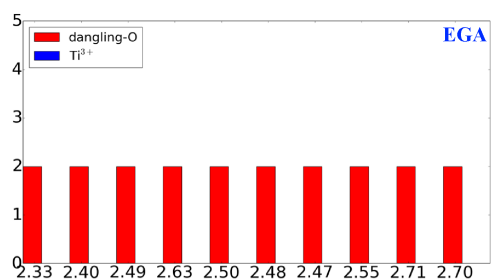

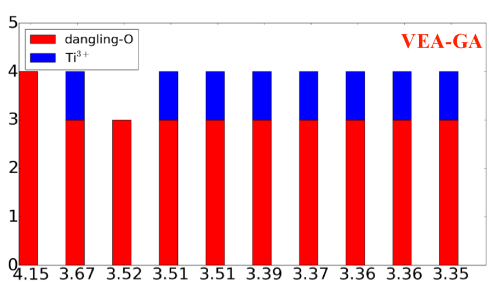

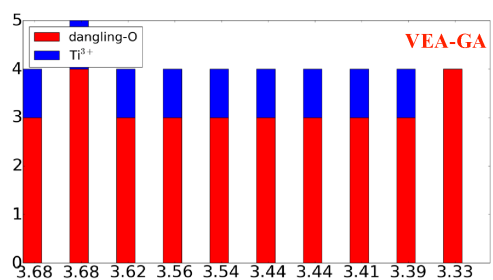

(e) $(TiO_2)_7$

(f) $(TiO_2)_8$

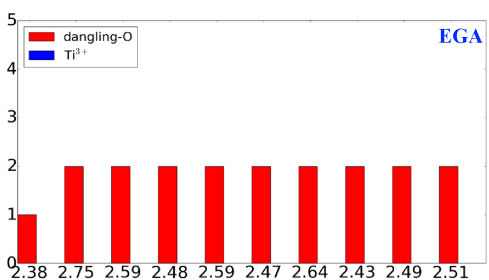

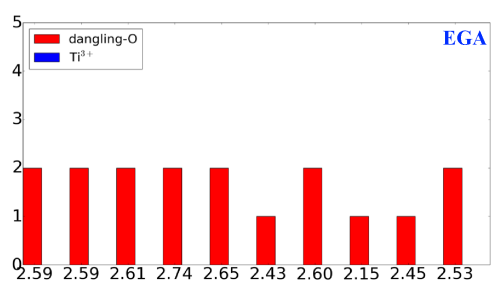

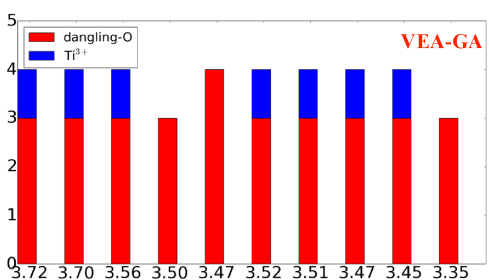

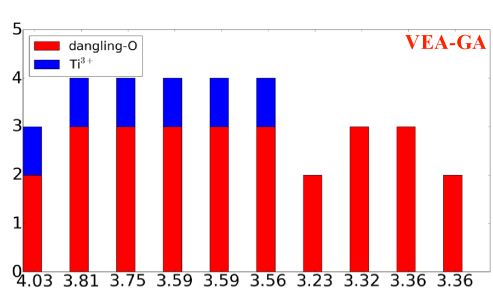

(g) $(TiO_2)_9$

(h) $(TiO_2)_{10}$

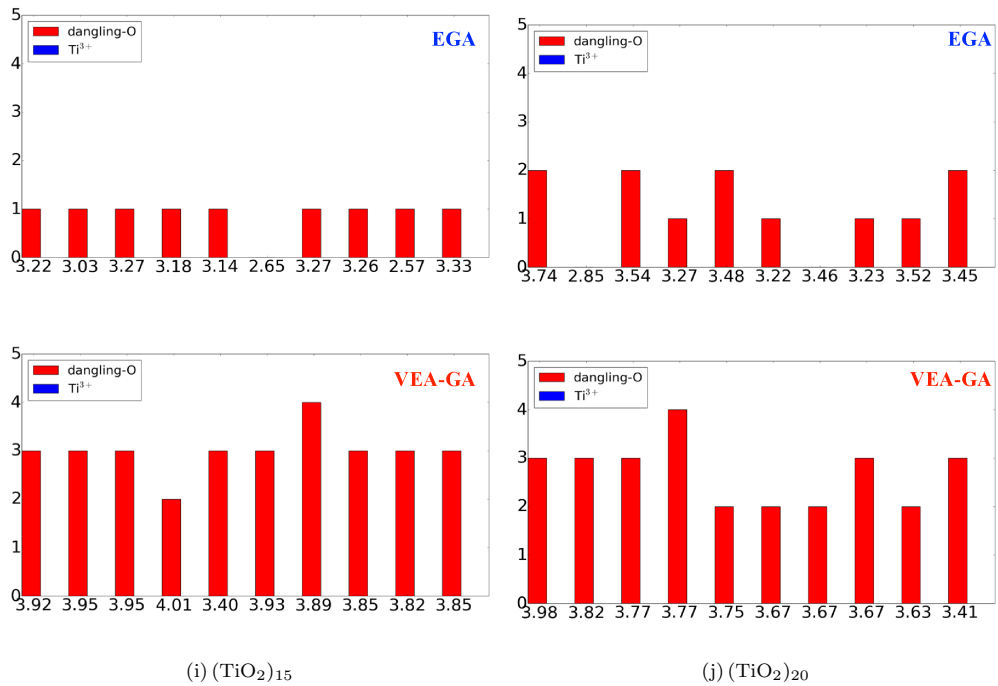

(i) (TiO$_2$)$_{15}$                                    (j) (TiO$_2$)$_{20}$

FIG. 4: We represent the structural features (namely, number of dangling-O, number of Ti$^{3+}$ site) in the best 10 isomers of EGA and VEA-GA. In the $x$-axis we label the VEA value of the isomer in eV. In the Y-axis the red bars represent the number of dangling-O and on top of them, the blue bars represent number of Ti$^{3+}$ site. Note that top of the red bar is the bottom of the blue bar. Thus if a red and blue bars end at 3 and 5 respectively, the number of dangling-Os are 3 and number of Ti$^{3+}$ sites are (5 - 3 =) 2.

### III. Ti-O PAIR CORRELATION FUNCTION (PCF) OF $(TiO_2)_n$ CLUSTERS:

We have plotted the Ti-O pair PCF of best 10 isomers (found by EGA and VEA-GA respectively) of the $(TiO_2)_n$ clusters. We have calculated all the Ti-O distances and divided them into 10 equally spaced bins to plot the histogram. The most significant difference in PCFs of the isomers found from VEA-GA and that of from EGA is the presence of the sharp peak (first peak in the PCFs as in Fig. 6) at 1.65 Å. This Ti-O bond length at 1.65 Å in the PCF represents the Ti-O bond of a dangling-O attached with a Ti-atom as shown in the Fig. 5 below. The intensity of this bond in PCF plot for VEA-GA-isomers is consistently higher than that of in EGA-isomers at all different sizes. This means all the isomers found by VEA-GA algorithm consists of more such dangling-Os in the structures.

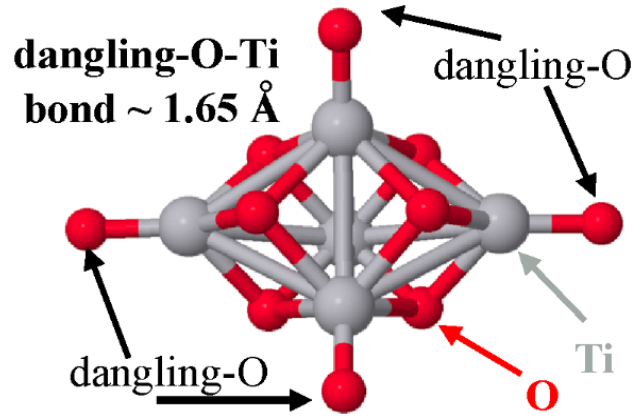

FIG. 5: Four dangling-Os in one VEA-GA isomer of $(TiO_2)_5$

# Ti-O PCF of $(TiO_2)_n$ clusters

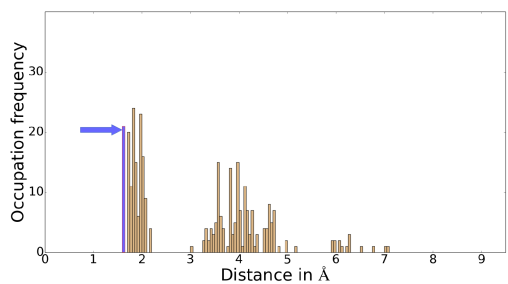

(a) EGA($(TiO_2)_4$)

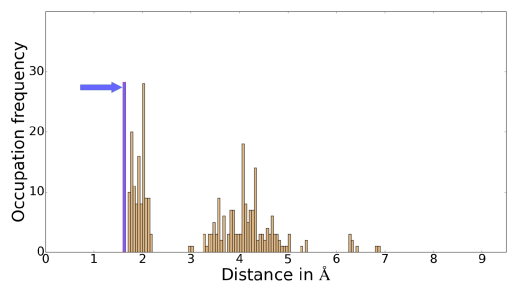

(b) VEA-GA($(TiO_2)_4$)

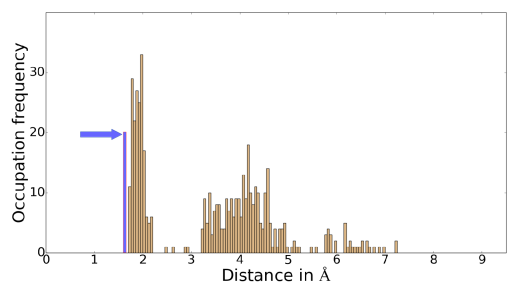

(c) EGA($(TiO_2)_5$)

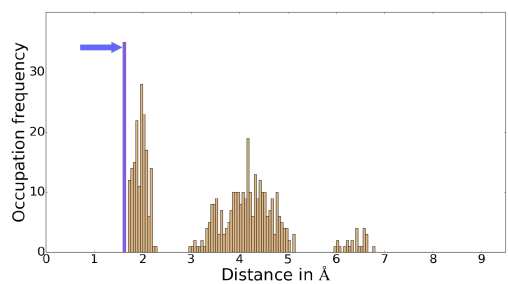

(d) VEA-GA($(TiO_2)_5$)

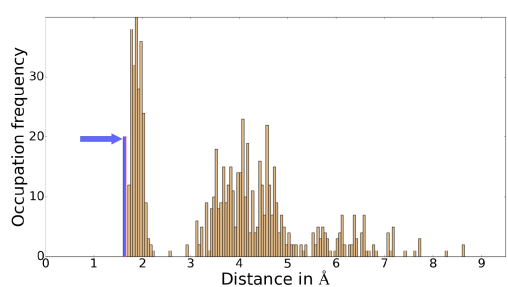

(e) EGA($(TiO_2)_6$)

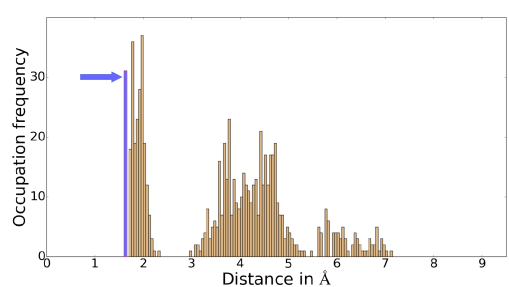

(f) VEA-GA($(TiO_2)_6$)

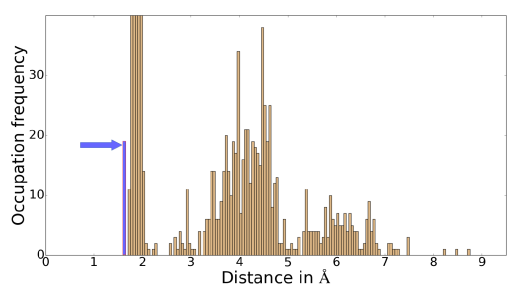

(g) EGA($(TiO_2)_7$)

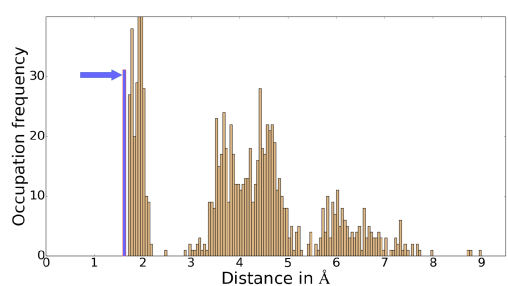

(h) VEA-GA($(TiO_2)_7$)

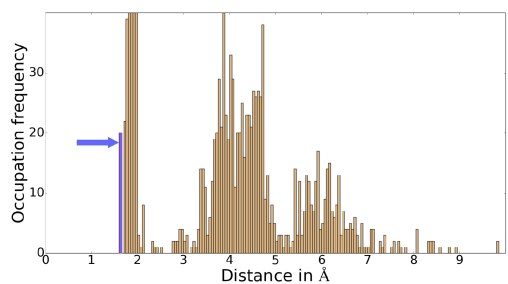

(i) EGA((TiO$_2$)$_8$)

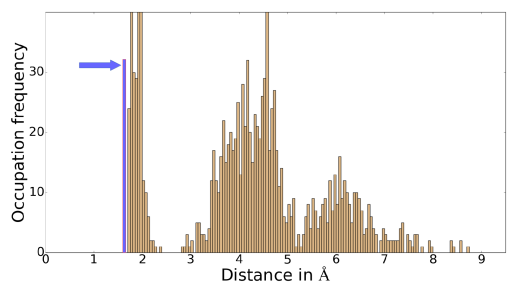

(j) VEA-GA((TiO$_2$)$_8$)

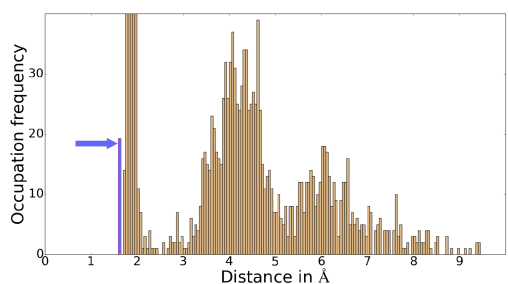

(k) EGA((TiO$_2$)$_9$)

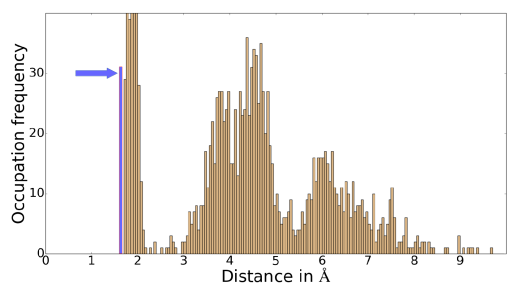

(l) VEA-GA((TiO$_2$)$_9$)

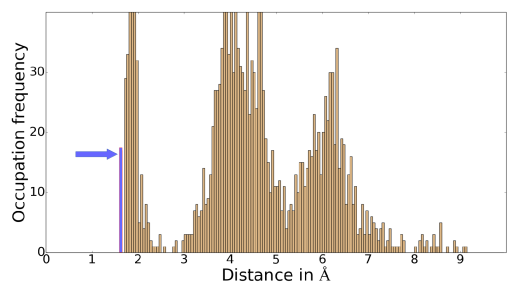

(m) EGA((TiO$_2$)$_{10}$)

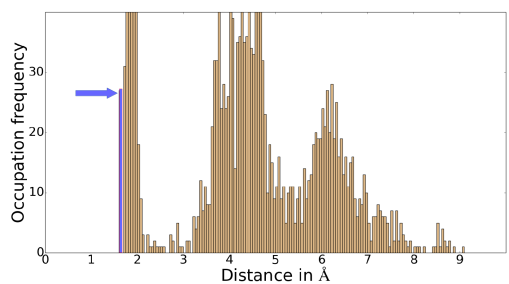

(n) VEA-GA((TiO$_2$)$_{10}$)

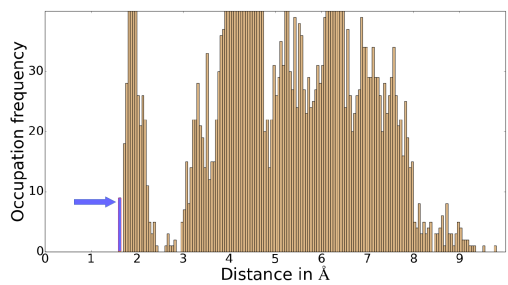

(o) EGA((TiO$_2$)$_{15}$)

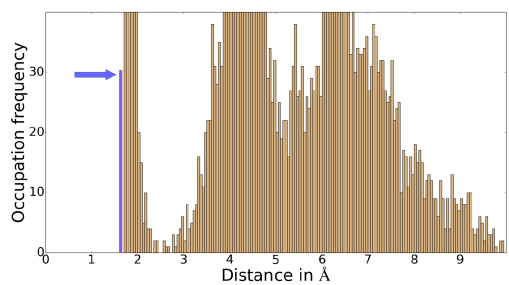

(p) VEA-GA((TiO$_2$)$_{15}$)

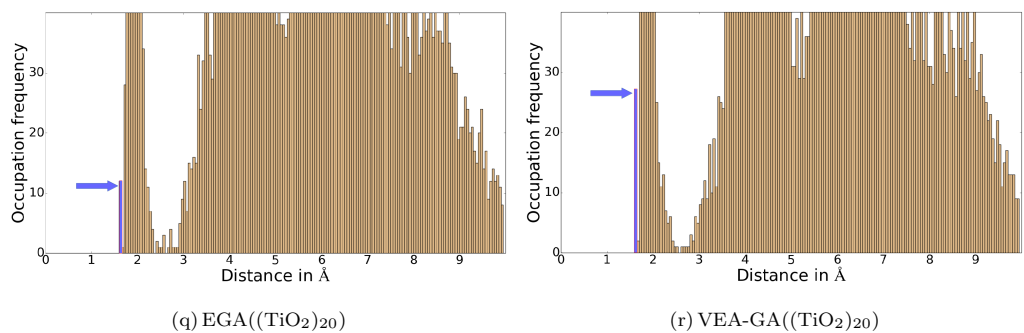

(q) EGA((TiO$_2$)$_{20}$)  (r) VEA-GA((TiO$_2$)$_{20}$)

FIG. 6: Ti-O PCF of (TiO$_2$)$_n$ clusters of the best 10 isomers found by EGA and VEA-GA. The striking difference is the presence of the sharp peak at 1.65 Å (the very first peak) in VEA-GA isomers. This peak is consistently higher in VEA-GA isomers as compared to EGA isomers at all studied sizes (e.g. n=4, 5,..., 10, 15, 20). This means all the VEA-GA isomers are having more Ti-O bonds of length 1.65 Å. This specific bond represents the dangling-O as shown in Fig. 5.

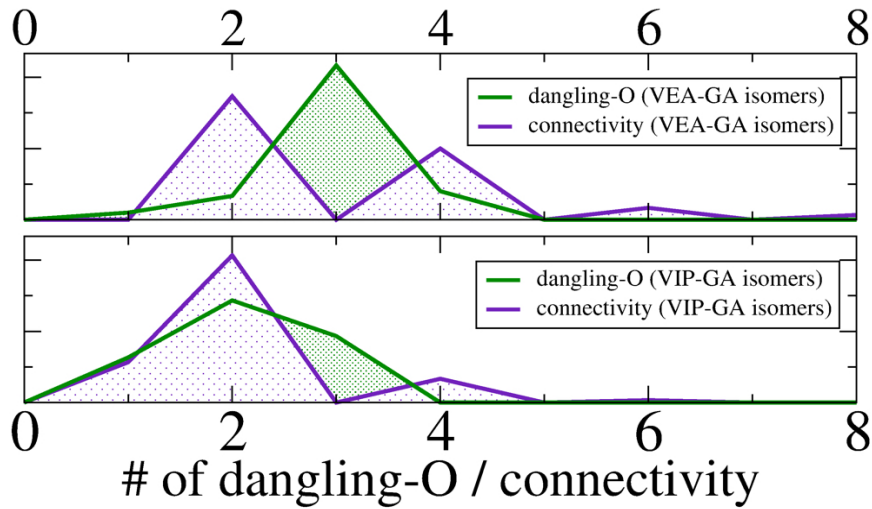

FIG. 7: Histograms of the abundance of isomers with a number of dangling-O atoms and low bond connectivity in the population of best isomers found by the VEA-GA and the VIP-GA.

## IV. LOW GAP ($\leq$ 5.5 eV) VS DANGLING-O AND CONNECTIVITY

Fig.7 shows histograms of the abundance of isomers with a number of dangling-O atoms and low bond connectivity in the population of best isomers found by the VEA-GA and the VIP-GA. The VEA-GA selects for structures with a large number of dangling-O atoms, regardless of their connectivity, whereas the VIP-GA selects for structures with a low connectivity regardless of the number of dangling-O atoms. We therefore would expect clusters with a narrow gap to have a number of dangling-O atoms, two of which are close to each other. Several of the structures with gaps below 5.5 eV, found by the VEA/VIP-GA, have 2-3 dangling-O atoms and a bond connectivity of two. We note that more structures with a narrow gap and more definitive structure-property relations may be found by tailoring a GA to search specifically for this property. We have found several structures at different sizes with gaps $\leq$ 5.5 eV. On inspecting their structure, we have found approximately 2-3 dangling-O atoms with a bond connectivity of two. We think that more structures with a narrow gap and more definitive structure-property relations may be found by tailoring a GA to search specifically for this property. Following Table depicts those details (i.e. no. of dangling-O, connectivity) of different isomers that are having gap $\leq$ 5.5 eV.

| Type of GA-scan | Gap value (eV) | Connectivity: | # of dangling-O |
|---|---|---|---|
| VIP-GA | | | |
| | 5.34361 | 2 | 2 |
| | 4.93132 | 2 | 3 |
| | 5.41224 | 2 | 2 |
| | 4.89849 | 2 | 2 |
| | 5.29608 | 2 | 2 |
| | 5.2959 | 2 | 2 |
| | 5.4691 | 2 | 2 |
| | 5.31938 | 2 | 2 |
| | 5.2723 | 2 | 3 |
| VEA-GA | | | |
| | 4.96373 | 4 | 2 |
| | 5.34427 | 2 | 3 |
| | 5.25586 | 6 | 2 |
| | 5.38608 | 4 | 2 |
| | 5.36256 | 2 | 3 |

## V. COORDINATES OF ALL ISOMERS AT ALL SIZES AND COMPARISON WITH PREVIOUSLY PUBLISHED STRUCTURES:

Our EGA was able to find all the low energy isomers reported previously [9, 14, 15] for all cluster sizes. For n=5 our choice of PBE0 as exchange-correlation functional and *higher-level* settings changes the ordering of the two most stable isomers, which are very close in energy, with respect to Ref. [9]. For n=10, we have found lower-energy structures than those previously reported. We have verified, at the PBE level, that the lowest energy, highest VEA, and lowest VIP isomers are stable minima in the sense of having no imaginary vibrational modes. All input and output files for the clusters described in this paper can be downloaded from `http://nomad-repository.eu`.

---